\documentstyle[preprint,aps,epsfig]{revtex}
\begin{document}
\newcommand{\be}{\begin{equation}}
\newcommand{\ee}{\end{equation}}
\newcommand{\beq}{\begin{eqnarray}}
\newcommand{\eeq}{\end{eqnarray}}
\newcommand{\ds}{\displaystyle}
\newcommand{\pia}{\mbox{$p_i^{\alpha}$}}
\newcommand{\pjb}{\mbox{$p_j^{\beta}$}}
\newcommand{\la}{\lambda_{\alpha}}
\newcommand{\bla}{\bar{\lambda}_{\alpha}}
\newcommand{\xa}{x^{\alpha}}
\newcommand{\ya}{y^{\alpha}}

\preprint{\rightline{IISc-CTS/9/98}}
\title{Autocatalytic Sets and the Growth of Complexity in
an Evolutionary Model}

\author{Sanjay Jain$^1$\footnote{email: \em{jain@cts.iisc.ernet.in}} and Sandeep Krishna$^2$
\footnote{email: \em{sandeep@physics.iisc.ernet.in}}\\
$^1${\it Centre for Theoretical Studies, 
Indian Institute of Science, Bangalore 560 012, India} \\
$^2${\it Department of Physics,
Indian Institute of Science, Bangalore 560 012, India }}
\maketitle
     
\begin{abstract} 
{ 
A model of $s$ interacting species is considered with two types of
dynamical variables. The fast variables are the populations of the
species and slow variables the links of a directed graph that defines
the catalytic interactions among them. The graph evolves via 
mutations of the least fit species. Starting from a sparse random
graph, we find that an autocatalytic set (ACS) inevitably appears
and triggers a cascade of exponentially increasing connectivity
until it spans the whole graph. The connectivity subsequently 
saturates in a statistical steady
state. The time scales for the appearance of an ACS in the 
graph and its growth have a power law dependence on 
$s$ and the catalytic probability. At the end of the growth period
the network is highly non-random, being localized on an exponentially small
region of graph space for large $s$.

} 
\end{abstract} \vskip 15pt

\newpage
A characteristic feature of 
chemical, biological, economic and social evolution
is that it produces a complex network of interactions among the 
component species or agents involved in it.  
Understanding the mechanisms responsible for the origin of such networks
and their movement towards greater complexity is an important issue.
One mechanism, based on quasi-species and  hypercycles \cite{ES}, proposed 
in the context
of prebiotic chemical evolution has self replicating entities as its 
components.
Another proposed mechanism starts from simpler components that are not
individually self replicating but can collectively form an ACS
\cite{Dyson}\cite{Kauffman}\cite{Wacht}. 
The present work attempts to explore the 
latter mechanism quantitatively through a mathematical model.  

The model has two main sources of inspiration. One is the
set of models studied by Farmer, Kauffman, Packard and others 
\cite{FKP}\cite{Kauffman} and by
Fontana and Buss \cite{FB}\ (see also \cite{SFM}\cite{SLKP}). 
Like these models the present one 
employs an artificial chemistry of catalysed reactions, albeit a
much simpler one, in which populations of species 
evolve over time. To this we add the feature, inspired by the
model of Bak and Sneppen \cite{BS}, that the least fit species
mutates. Unlike the Bak-Sneppen model however,
the mutation of a species also changes its links to other
species. This allows us to investigate 
how the network of interactions among the species evolves over
a longer time scale.
We find that for a fixed total number of species, the network
inevitably evolves towards a higher complexity as measured by
the degree of interaction among species and their interdependence. The
increase is triggered by the chance appearance of an ACS, is exponential
in time, and leads to a highly non-random organization. 

The system is described by a directed
graph with $s$ nodes. The nodes represent the components or species and
the directed links represent catalytic interactions among them. 
A link from node $j$ to $i$ means that species 
$j$ catalyses the production of $i$.
The graph is completely described by specifying
the adjacency matrix $C \equiv (c_{ij})$, $i,j =1, \ldots , s$.
$c_{ij}$ equals unity if there is a link from $j$ to $i$, and zero
otherwise. A link from a node to itself is prohibited
(diagonal entries of $C$ are zero);
this corresponds to the exclusion of self-replicating species.

At the initial time ($n=0$), the graph is random. That is,
$c_{ij}$ (for $i \neq j$)
is unity with probability $p$ and zero with
probability $1-p$. Thus on average every row and column of $C$
has $m \equiv p(s-1)$ non-zero entries, representing the average
number of links to and from a node. $p$ is the probability
that a given species will be a catalyst for another given species.

The graph is updated at discrete time steps
($n=1,2,\ldots$) as follows: One 
selects the `mutating node'
of the existing graph by a rule
to be specified below, removes all the existing incoming
and outgoing links
to and from that node, and replaces them by randomly chosen links
with the same catalytic probability $p$.
That is, if the selected node is $i$, the $i^{th}$ row
and column of $C$ are reconstituted according to the same rule as in
the previous paragraph. This changes the graph into a new one.
A mutating node is selected afresh and 
this process is iterated over many time steps.

The mutating node at any $n$ is taken to be the one with
the `least fitness' at that time step. 
Associated with every node $i$ is a
population $y_i \geq 0$, or a relative population 
$x_i \equiv y_i/Y$, $Y \equiv \sum_{j=1}^s y_j$.
The population depends upon a continuous
time $t$, and its evolution between two successive graph updates
(i.e., while the graph $C$ remains fixed) is given by
\be
\dot{y}_i = \sum_{j=1}^s c_{ij} y_j - \phi y_i.
\label{ydot}
\ee
From this it follows that $x_i$ has the
dynamics
\be
\dot{x}_i = \sum_{j=1}^s c_{ij} x_j -  x_i \sum_{k,j=1}^s c_{kj} x_j.
\label{xdot}
\ee
The $x_i$ dynamics depends only on $C$ and not on $\phi$. The
time between two successive graph updates is assumed long enough to allow
the fast variables $x_i$ to reach their attractor configuration,
denoted $X_i$. $X_i$ 
is a measure of the fitness of the species $i$ in the environment
defined by the graph. The set of nodes with the smallest value of 
$X_i$ is called the set of least fit nodes. The mutating
node is picked randomly from the set of least fit nodes.  
For the cases that have arisen in our simulations, the set of least fit 
nodes depends only on the graph and not upon the initial values of 
$x_i$. 

The dynamics (\ref{ydot}) is an approximation of the rate equations
in a well stirred chemical reactor with a nonequilibrium dilution flux $\phi$
when the reactants necessary for the production
of the molecular species in the graph are fixed and in abundance, and 
spontaneous
reaction rates are much smaller than catalysed reaction rates. 
Then reaction rate is limited by and proportional to
catalyst concentration.  
The $i^{th}$ species grows via the catalytic action of all species $j$
that catalyse its production and declines via a common death rate 
$\phi$. That all catalytic strengths 
are equal is an idealization of the model. This dynamics might
also be relevant for economics (e.g., positive feedback
networks) as well as certain kinds of ecological webs. A justification
for selecting the least fit node to be mutated in a 
molecular context is that the species with the least population is the
most likely to be lost in a fluctuation
in a hostile environment. Alternatively, in an
ecological context, certain fitness landscapes might be such that a
low fitness is correlated with 
a smaller barrier to mutation (see the arguments in \cite{BS}).
In economics a correlation between fitness and survival
is at the heart of evolutionary game theoretic models like the
replicator equation; the elimination of the least fit is an
extreme idealized case of this correlation.
Keeping the total number
of species constant in the simulation is another idealization of
the model. 

In fig. 1 we plot the total number of links $l(n)$ in the graph
versus the graph update time step $n$. Three runs with $s=100$
are exhibited, each with a different value of $p$ (or $m$).
We have studied the parameter range
$m$ from $0.05$ to $2.0$ and $s=50,100,150$.
For fixed $m,s$
we have conducted several runs with different
random number seeds. The runs shown in fig. 1 
are typical of the runs with the same parameter values.
 
The curves have three distinct regions. Initially the number of links
hovers around the value expected for a random graph, $l \sim ms$.
The second region is one of rapid increase, in which $l$ rises several fold.
The third is a statistical steady state with
many fluctuations where the average connectivity is much higher than
the initial one.

The increase in $l$ over time is a consequence of selection.
In a `random run' in which
the mutating node is chosen at random from among all the
$s$ nodes instead of from the set of least fit nodes, 
$l$ keeps fluctuating about its initial average value $ms$.
Note that under selection $l$ rises even though the average
connectivity of the species that replace the mutating species
is the same as in the initial random graph.
One may be tempted to give the following naive explanation of the increase:
From (\ref{ydot}) it is clear that the larger the number of
species that have links
to the species $i$, the greater is the rate of increase
of $y_i$. Therefore the species that
do well (population wise) are those that have more links coming in,
and conversely
those that don't do so well are deficient in incoming links. Hence
selecting the least fit species amounts to selecting the species
that have lower connectivity than average. If these are
replaced by species that have the old average connectivity, it
is no surprise that the number of links increases.

This explanation is not correct. Firstly, this argument is unable
to explain the observed fact that there is a long region of
almost constant $l$ in the graphs before it starts to increase.
Second, in this region the mutating nodes tend to
have a larger proportion of outgoing links than average, which
more or less balances out their deficiency in incoming links.
The real explanation, which we substantiate in detail below, is
the chance appearance of an ACS in the graph. 

Since (\ref{xdot}) does not depend on $\phi$, we can set $\phi = 0$
in (\ref{ydot}) without loss of generality for studying the attractors of
(\ref{xdot}). For fixed $C$
the general solution of (\ref{ydot}) is 
${\bf y}(t) = e^{Ct}{\bf y}(0)$, where ${\bf y}$ denotes
the $s$ dimensional column vector of populations.
It is evident that if ${\bf y}^{\lambda} \equiv (y_1^\lambda, \ldots,
y_s^\lambda)$ viewed as a column vector
is a right eigenvector of $C$ with eigenvalue $\lambda$,
then ${\bf x}^{\lambda} \equiv 
{\bf y}^{\lambda}/\sum_i^s y_i^\lambda$ is a fixed point of (\ref{xdot}).
Let $\lambda_1$ denote the eigenvalue of $C$ which has the
largest real part; it is clear that ${\bf x}^{\lambda_1}$ is an
attractor of (\ref{xdot}). 
By the theorem of Perron-Frobenius for non-negative matrices \cite{MM}\
$\lambda_1$ is real
and $\geq 0$ and there exists an eigenvector
${\bf x}^{\lambda_1}$ with $x_i \geq 0$. 
If $\lambda_1$ is nondegenerate, $\bf{x}^{\lambda_1}$ is the
unique asymptotically stable attractor of (\ref{xdot}),
${\bf x}^{\lambda_1} = (X_1, \ldots, X_s)$.
In practice, we found in
our simulations that $\lambda_1$ was usually nondegenerate, except
for very sparse graphs. This is not surprising in view of
the well known level repulsion in random matrix theory, which
implies that repeated eigenvalues are very improbable for a generic
matrix.

An ACS is defined as a subgraph whose every node
has at least one incoming link from a node that belongs to the
same subgraph. This definition is meant to capture the property
that an ACS has `catalytic closure'
\cite{Kauffman}, i.e., it contains the catalysts for all its members. 
The simplest ACS
is a 2-cycle. The following hold:
(i) An ACS always contains a cycle. (ii) If a graph has no ACS
then $\lambda_1 = 0$ for the graph. (iii) If a graph has an ACS
then $\lambda_1 \geq 1$. (iv) If $\lambda_1 \geq 1$, then the subgraph
corresponding to the set of nodes $i$ for which $x_i^{\lambda_1} > 0$
is an ACS. We will call this subgraph the `dominant ACS' of the graph.
These properties, which we first observed numerically, can be proven 
analytically from graph theory \cite{BH}.
It follows from (iv) that members of the dominant ACS completely
overshadow all other species population wise.

For $m < 1$ the initial random graph
is sparse. E.g., with $s=100$, $m=0.25$, there are on average
only 25 links. Most of the nodes are singletons, some pairs have
a single link among them, and there are a few chains or other
trees with two or more links. 
The probability of there being a cycle is small ($ \sim O(m^2)$).
Fig. 2 shows how $\lambda_1$ evolves.
Since $\lambda_1$ remains zero for $n < n_1 = 1643$ it is clear that
there is no cycle in the graph in this period. 
When there are no cycles, then $y_i \sim t^r$ for large $t$, where
$r$ is the length of the longest path terminating at $i$. Then
$X_i = 0$ for all $i$ except 
the nodes at which the longest paths in the graph terminate.
Define $s_1(n)$ as the number of species $i$ for which 
$X_i \neq 0$ at the $n^{th}$ time step. This is plotted
in fig. 3. (The lower curve in figs. 2 and 3 
corresponds to a `random run' with $s=100,m=0.25$.)
Since the mutant can be any least fit node, the chains can be disrupted
over time. In particular if the mutating node happens to be the `nearest
neighbour' of a node that is the terminating point of the longest chain,
after mutation the latter node can become a singleton and join the
ranks of the least fit.

The picture changes qualitatively when an ACS appears by chance
in a mutation. Then $\lambda_1$ jumps from zero to one (at $n=n_1$).
For concreteness lets say the ACS at $n=n_1$ is a 2-cycle between species 1
and 2. Then ${\bf x}^{\lambda_1}= (1/2, 1/2, 0, 0, \ldots, 0)$. The key
point is that both species 1 and 2 are absent from the set of least
fit nodes and will not be mutated at the next time step. 
By definition the nodes which are
not part of the dominant ACS of a graph with $\lambda_1 \geq 1$
have $x_i^{\lambda_1} = 0$ from property (iv) above, and hence
constitute the set of least fit nodes. Therefore,
as long as the dominant ACS does not include the
whole graph, the mutating node will be outside it, and
hence a mutation cannot destroy the links that make up the dominant ACS
just before the mutation.
{\it Thus the
auto-catalytic property is guaranteed to be preserved 
once an ACS appears until the dominant ACS engulfs the whole graph.}
In the run of figs. 2 and 3 this happens at $n=n_2=2589$, when $s_1(n) = s$.
Whenever $s_1 < s$, $\lambda_1$ is a non-decreasing function
of $n$. An increase in $s_1$ during $n \in (n_1, n_2)$ occurs whenever
a mutant species gets an incoming link from the existing dominant ACS
and hence becomes a part of it. (Note that $s_1$ itself need not be
a non-decreasing function of $n$ when it is $< s$, because the 
{\it dominant} ACS after a mutation can be smaller than the one before the
mutation.)  

There is another qualitative change in the evolution at $n=n_2$. Since the
whole graph becomes an ACS, for the
first time since the appearance of the ACS the mutant must now
be from the dominant ACS itself. When the mutant happens to be
a species which is
playing an important catalytic role in the organization (a `keystone
species'), the mutation can disconnect a number of other species from
the main ACS, as evidenced from the substantial drop
in $s_1$ at $n=4910$. 
The final steady state in fig. 1 is characterized by the fact that the
mutating node has, on average, the same total number of links (namely,
$2m$) as its replacement. 

The above picture holds for different values of $m,s$, as long as 
$m$ is small enough and $s$ large enough. 
For very small $m$ the fluctuations in the final steady state are
large; the ACS can even be destroyed completely.
For sufficiently
high values of $m$ the initial random graph is dense enough to contain 
an ACS, hence the initial period with $\lambda_1=0$ is
absent. 

During the growth period $n \in (n_1,n_2)$, $s_1$ and $l$
(locally averaged in time) 
grow exponentially. E.g., 
$s_1(n) \sim s_1(n_1) e^{(n-n_1)/\tau_g}$. The $m$ dependence of 
the `growth time scale' $\tau_g$ is shown in fig. 4 and is consistent
with $\tau_g \propto m^{-1}$. In a time $\Delta n$, the average increase in 
$s_1$ in a large sparse graph is given by $\Delta s_1 \sim 
ps_1 \Delta n$, which is the average number of mutating nodes 
out of $\Delta n$
which will get an incoming link from the $s_1$ nodes of the dominant ACS.
Therefore $\tau_g \sim 1/p \sim s/m$. The average `time of arrival'
$\tau_a \equiv \langle n_1 \rangle$ of an ACS in a sparse graph
is given by $\tau_a \sim s/m^2 \sim 1/(p^2s)$, since the probability
that a graph which does not have an ACS will get a 2-cycle at the next
time step (3-cycles and larger ACSs being much less likely for small $p$)
is $\sim p^2s$. Thus for any finite $p$, however small, the 
appearance and growth of ACSs in this model is inevitable.

The graphs generated at the end of the growth phase are highly non-random. 
The probability of a random graph with $s$ nodes and on average $m^{*}$
links per node being an ACS is $[1-(1-[m^{*}/(s-1)])^{s-1}]^s$,
which declines exponentially with $s$ when $m^{*} \sim O(1)$. 
This may be relevant to the
origin of life problem for which naive estimates of the probability of 
a complex organization like a cell arising by pure chance on the prebiotic
earth give exponentially small values.
The present model provides an example whereby highly non-random 
organizations can arise in a short time by a mechanism that causes an
exponential increase in complexity. The hypercycle is known to
suffer from the short-circuit instability which reduces its complexity
\cite{NHB}.
It is interesting that in the present model ACSs provide the system with
the opposite kind of instability, in the direction of increasing complexity.
Finally this model provides an example of how selection for fitness at
the level of individual species results, over a long time scale, in 
increased complexity of interaction of the collection
of species as a whole. Note in fig. 2 that in the random run ACSs come
and go, whereas, when selection is operative, the system `cashes in'
upon the novelty provided by an ACS that arises by chance. This is 
reminiscent of how ecosystems and economic webs capitalize on 
`favourable' chance events.

We thank J. M. Berg, A. Bhaduri,
V. S. Borkar, C. Dasgupta, R. Hariharan, G. Rangarajan and B. S. Shastry
for discussions. S. J. acknowledges the affiliation and
support of the Jawaharlal Nehru Centre for Advanced Scientific Research
(JNCASR), Bangalore, as well as Associate membership 
and hospitality of the Abdus Salam International
Centre for Theoretical Physics, Trieste. He also thanks the School of Physical
Sciences, Jawaharlal Nehru University, New Delhi, for providing necessary 
facilities while this work was in progress. S. K. thanks JNCASR for a 
Summer Research Fellowship during 1996 and 1997 during which this work
was begun.

\pagebreak
\begin{figure}
\centerline{\epsfbox{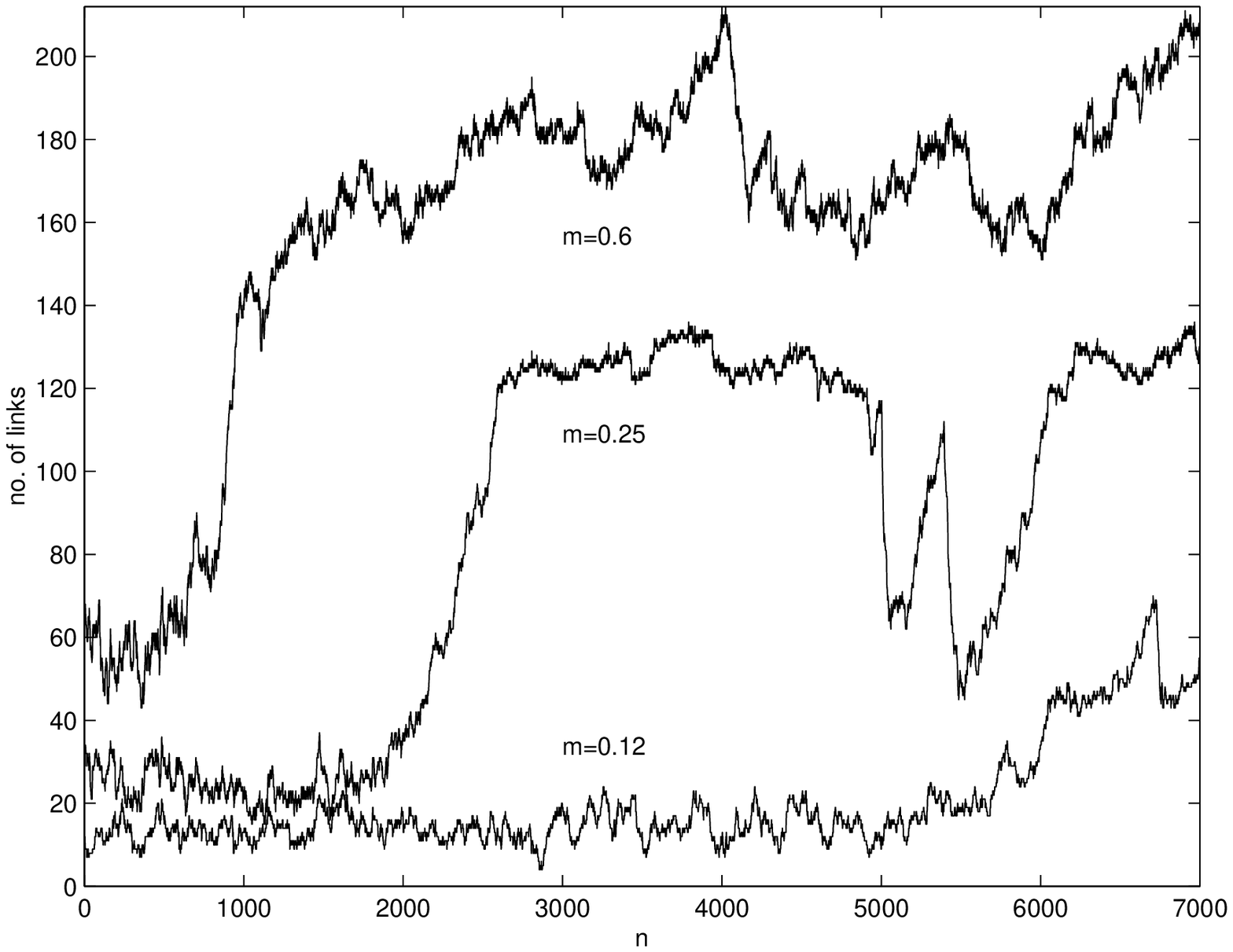}}
\narrowtext{\caption{Total links versus $n$ for three runs with $s=100$.}}
\end{figure}

\pagebreak
\begin{figure}
\centerline{\epsfbox{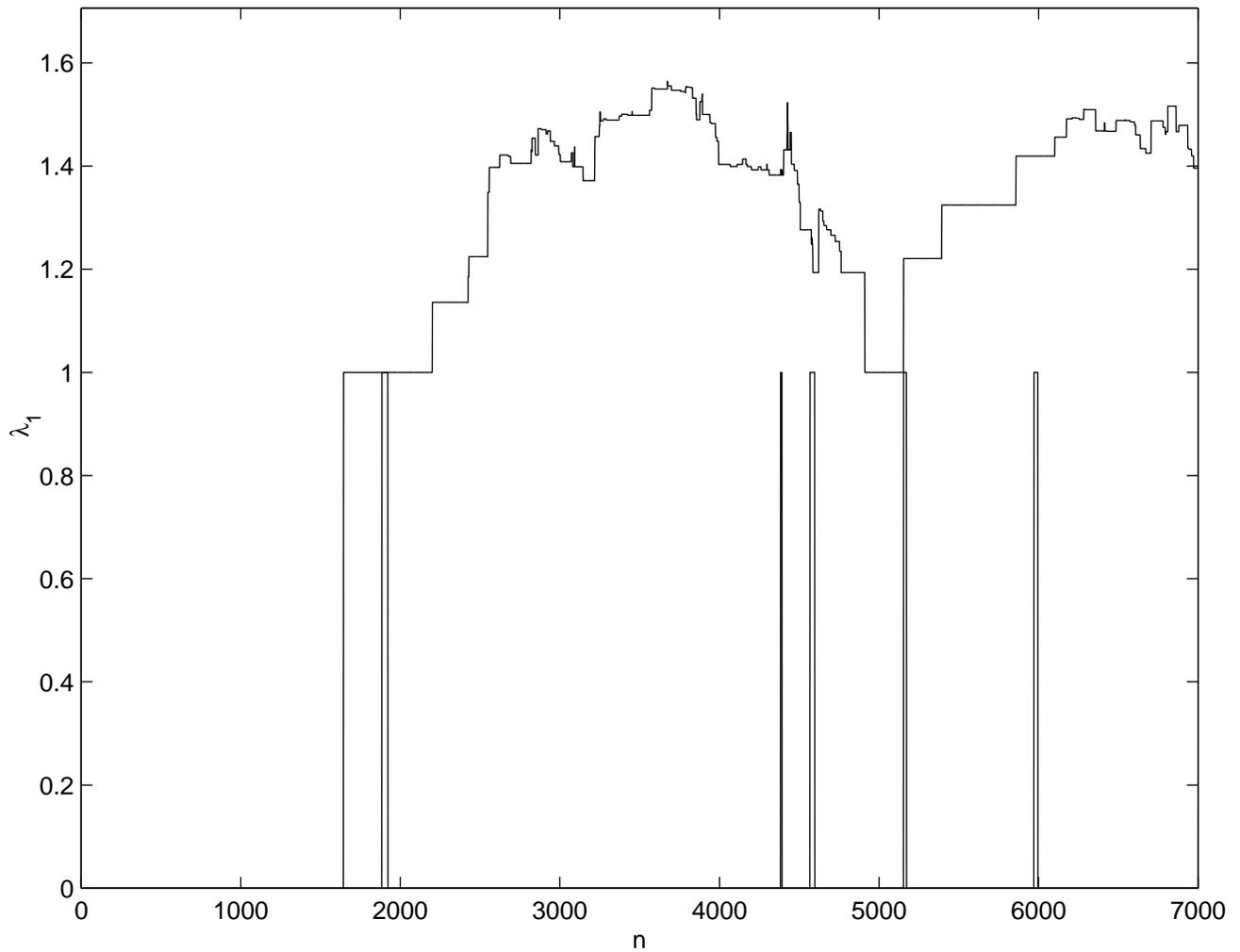}}
\narrowtext{\caption{$\lambda_1$ versus $n$ for $s=100, m=0.25$. The
upper curve is for the same run as the middle curve of fig. 1.
The lower curve is for a random run with $s=100, m=0.25$.}}
\end{figure}

\pagebreak
\begin{figure}
\centerline{\epsfbox{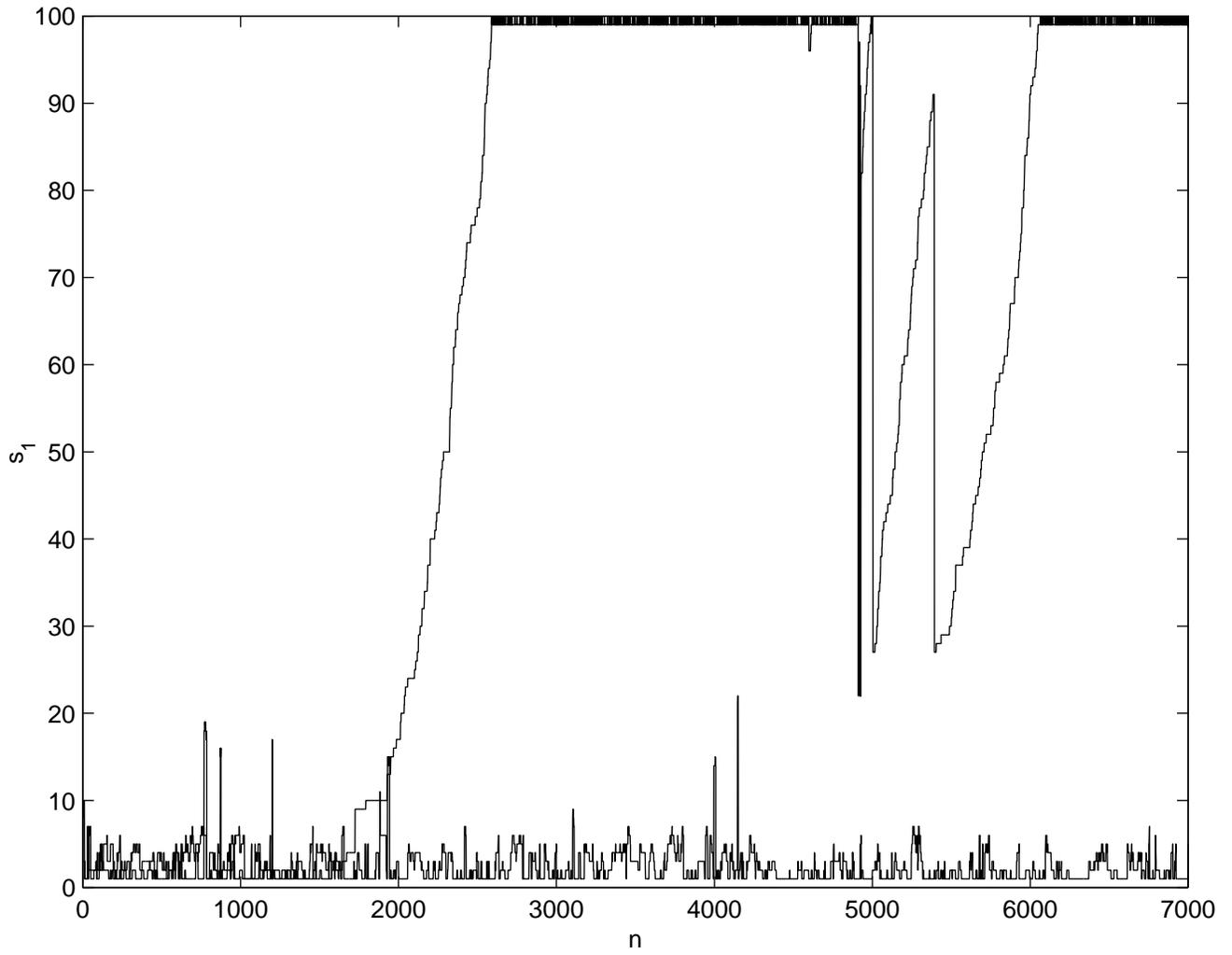}}
\narrowtext{\caption{$s_1$ versus $n$ for the same runs as in fig. 2.}} 
\end{figure}

\pagebreak
\begin{figure}
\centerline{\epsfbox{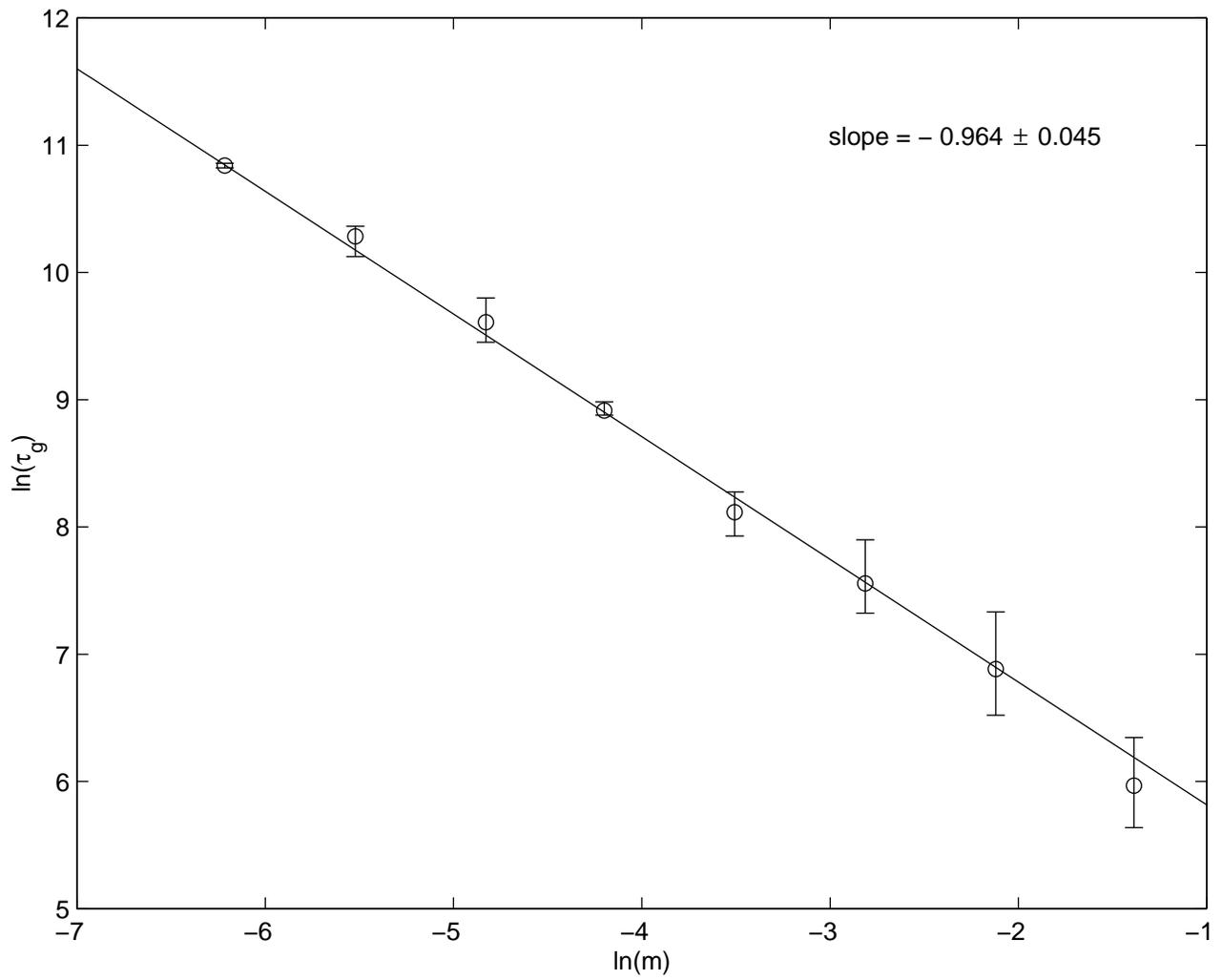}}
\narrowtext{\caption{Power law dependence of $\tau_g$ on $m$.}} 
\end{figure}

\end{document}